\documentclass[%
reprint,
notitlepage,
 aip,
jcp,
 longbibliography
]{revtex4-1}

\usepackage{amsmath}
\usepackage{amsthm}
\usepackage{dsfont}
\usepackage{mathbbol}
\usepackage{bbold}
\usepackage{mathrsfs}
\usepackage{graphicx}% Include figure files
\usepackage{dcolumn}% Align table columns on decimal point
\usepackage{bm}% bold math
\usepackage{comment}% comment text
\usepackage{verbatim}% comment text
\usepackage[usenames]{color}
\usepackage[normalem]{ulem}
\usepackage{cancel} %cross out text obliquely

\usepackage[breaklinks]{hyperref}

\hypersetup{
    unicode=true,
    colorlinks=true,
    linkcolor=blue,
    citecolor=blue,
    urlcolor=blue
  }
\usepackage{breakurl}

\usepackage[per-mode=symbol]{siunitx}   
\usepackage[version=3]{mhchem}
 
\DeclareSIUnit\intensity{\watt\per\centi\meter\squared}
\DeclareSIUnit\fieldstrength{\volt\per\centi\meter}
\DeclareSIUnit\kfieldstrength{k\volt\per\centi\meter}
\DeclareSIUnit\energy{cm^{-1}}

\newcommand{\melement}[3]{\ensuremath{\left\langle #1 \left|#2\right|#3\right\rangle}}%
\newcommand{\ie}{i.\,e.}%
\newcommand{\expected}[1]{\left\langle #1\right\rangle}
\newcommand{\granada}{\affiliation{Instituto Carlos I de F\'{\i}sica Te\'orica y Computacional, and Departamento de F\'{\i}sica At\'omica, Molecular y Nuclear, Universidad de Granada, 18071 Granada, Spain}}%

\newcommand{\ucm}{\affiliation{Departamento de Qu\'imica F\'isica, Universidad Complutense de Madrid, 28040 Madrid, Spain}}

\begin{document}

\title{Full Control of non-symmetric molecules orientation using weak and moderate electric fields}
\author{Rosario González-Férez}\granada\author{Juan J. Omiste}\email{jomiste@ucm.es}\ucm
\date{\today}
\begin{abstract}
We investigate the full control over the orientation of a non-symmetric molecule by using moderate and weak electric fields. 
Quantum Optimal Control techniques allow us to orient any axis of 6-chloropyridazine-3-carbonitrile,
which is taken as prototype example here, along the electric field direction.
%lanar molecule lacking rotational symmetry. 
We perform a detailed analysis by exploring the impact on the molecular orientation of the time scale and strength of the control field.
The underlying physical phenomena allowing for the control of the orientation  are interpreted in terms of the frequencies contributing 
to the field-dressed dynamics and to the driving field by a spectral analysis. 
\end{abstract}

\maketitle

\section{Introduction}
\label{sec:introduction}

%The ability to control the orientation of molecular axes is a crucial task for a range of applications, from chemical reaction selectivity to quantum information processing. Optimal control algorithms have been proposed to achieve this goal~\cite{Werschnik2007}, and the Krotov algorithm has been shown to be particularly effective~\cite{Ohtsuki2004}.  In this paper, we use Quantum Optimal Control techniques to investigate the full control over the orientation of non-symmetric molecules using weak and moderate electric fields. Specifically, we explore the impact of various time scales and strengths of the control field on the orientation of 6-chloropyridazine-3-carbonitrile, a planar molecule lacking rotational symmetry. Additionally, we conduct a spectral analysis of the control field to gain a deeper understanding of the underlying physical phenomena contributing to the control. Our results contribute to the ongoing effort to achieve full control over the orientation of molecular axes, with potential applications in a range of fields, from chemistry to quantum computing.

The interaction of external fields with molecules
%control and manipulate their rotational dynamic
creates coherent superpositions of field-free rotational states~\cite{stapelfeldt:rev_mod_phys_75_543,Koch2019} 
with plethora of applications in a wide range of areas
from stereodynamics~\cite{herschbach:epjd38,aquilanti:pccp_7,Yang2022}, high precision measurements~\cite{Kozyryev2017},
to quantum information processing~\cite{DeMille2002a,Albert2020}.
%%%% RGF yo solo podnría reviews en ests primera referencia y algun articulo de Bretislav y Herschbach, que no los citamos y son los pioneros 
%The cs of  molecules with external fields 
%creates a coherent superposition of field-free rotational states, giving rise
%orientation of asymmetric molecules is of
%is of special interests due to its possible applications in a wide range of fields. 
%, from chemical reaction selectivity to quantum information processing. 
%either to the alignment~\cite{stapelfeldt:rev_mod_phys_75_543,nevo:pccp11,Leibscher2003}
%or orientation~\cite{loesch:jcp93}.
%The applications of these phenomena 
 %Among the most prominent applications of these techniques, let us remark 
 This interaction allows for the creation and control of highly rotating molecular states, namely 
 superrotors~\cite{Korobenko2014, Korobenko2015a,Milner2016a}, the separation and manipulation of chiral 
 molecules~\cite{Tutunnikov2021a,Patterson2013,Leibscher2022a,Sun2023}, 
 the orientation of proteins~\cite{Marklund2017,Sinelnikova2021}, 
 the isotopic separation~\cite{Kurosaki2022}, the controlled manipulation enantiomers of chiral molecules~\cite{Saribal2021},

The orientation of the molecule requires the angular confinement of a
molecule-fixed axis along a laboratory-fixed one and a preferred direction of 
the electric dipole moment~\cite{loesch:jcp93,rost:prl68,friedrich:jcp111}.
For asymmetric molecules, the 3D orientation implies that 
the all three molecular axes of inertia are confined to laboratory fixed ones and a well defined 
direction of electric dipole moment~\cite{nevo:pccp11,Kienitz2017}. 
 Certain complex molecular systems, such as, the planar molecule 6-chloropyridazine-3-carbonitrile and 
 the chiral one bromochlorofluoromethane (CHBrClF),  possess a permanent dipole moment which do not coincide with any inertia axis.  For these systems,  
 it may be necessary to control the orientation of a specific molecular
 axis different from the one defined by the electric dipole moment.
 Therefore, it is mandatory to design an specific protocol beyond brute force orientation~\cite{bulthuis:jpca101} or standard mixed-field orientation~\cite{loesch:jcp93,friedrich:jcp111} to accurately restrict the position of a given axis, \ie,  a given atom.
Such a task could benefit from long-established algorithms, as the Quantum Optimal Control (QOC)~\cite{Tannor1986},
to accurately drive the molecular dynamics.
The fundamental concept of QOC involves designing a driving 
field that optimizes a certain expectation value or a state population in a quantum system.
This is done by solving a self-consistent set of equations, 
for which several algorithms have been developed~\cite{Werschnik2007,Koch2016}. 
QOC constitutes a powerful tool for manipulating the dynamics of quantum systems~\cite{Tannor1986,Kosloff1989,Koch2016},
including the control of qubits~\cite{Ansel2022}, population of vibrational states~\cite{Delgado-Granados2022}, 
coupled spin dynamics~\cite{Khaneja2005}, among others.
Regarding molecular orientation control, Krotov algorithm~\cite{Krotov1983, Tannor1992,Werschnik2007} is used to design 
terahertz laser pulses~\cite{Dmitriev2023}, or to manipulate the permanent dipole orientation of molecular ensembles, 
as reported in various studies~\cite{Coudert2017, Beer2022, Damari2022, Coudert2018}. In addition, this algorithm has been 
used to optimize alignment 
of molecules immerse in a thermal bath~\cite{Pelzer2008,Salomon2005}. 
 
 In this work, we consider a non-symmetric molecule, and apply the QOC algorithm to optimize the orientation of any molecular axis.
 Our study provides physical insights into the design of driving fields for this purpose. 
In particular, we provide a deeper understanding of the mechanisms underlying the control by means of a spectral 
 analysis of the field-dressed wavepacket and of the driving field.
This paper is organized as follows.~\autoref{sec:theory_and_methods} provides the Hamiltonian in an electric field, and
a brief overview of the QOC theory with the corresponding equations of motion.
In \autoref{sec:results}, we present the QOC results
for the planar molecule 6-chloropyridazine-3-carbonitrile (CPC), taken here as a prototype example,
orienting 
 its permanent dipole moment and principal axes of inertia within the molecular plane, while varying the control field strength and duration. Finally, \autoref{sec:conclusions_and_outlook} summarizes the main findings of the study and presents a short outlook.

\section{Theory and Methods}
\label{sec:theory_and_methods}

\subsection{The Hamiltonian}
\label{sec:molecular_system}

In this work we investigate the orientation of a planar molecule without rotational symmetry by means of an external
time-dependent electric field $\vec{E}(t)$, which is taken parallel to the Laboratory Fixed Frame (LFF) $Z$-axis. 
We work within the Born-Oppenheimer and the rigid rotor approximation, therefore, the molecular system is described by the Hamiltonian
\begin{equation}
    \label{eq:hamiltonian}
    H(t)=H_{rot}+H_{E}(t),
\end{equation}
where $H_{rot}=B_{x}J_x^2+B_{y}J_y^2+B_{z}J_z^2$ is the field-free Hamiltonian, $J_k$ the projection of the angular momentum operator $\vec{J}$ along the $k$ axis of the molecular fixed frame (MFF). 
The field-free rotational states are denoted by $J_{K_a, K_c}M$~\cite{Omiste2011}, 
where $J$ is the total angular momentum number and $M$ the magnetic quantum number, \ie, 
the projection of $\vec{J}$ along the LFF $Z$-axis,  
and $K_a$ and $K_c$ are the projection of $\Vec{J}$ along the MFF $z$-axis in the prolate  and oblate  limiting cases, respectively.

The coupling of the molecular
electric dipole moment $\vec{\mu}$~with an electric field
parallel to the LFF $Z$-axis~\cite{bulthuis:jpca101,Kong:jpca104}
reads 
\begin{eqnarray}
    \nonumber
    H_E(t)&=&-\vec{\mu}\cdot\vec{E}(t)=-E(t)\mu\cos\theta_{Z\mu}=\\
    \label{eq:efield}
    &=& -E(t)\left(\mu_z\cos\theta_{Zz}+\mu_x\cos\theta_{Zx}\right)
\end{eqnarray}
with being $\theta_{Z\mu}$ the angle between $\vec{\mu}$ and the LFF $Z$-axis, 
and $\theta_{Zk}$ the Euler angle between the molecular $k$ axis and LFF $Z$-axis~\cite{Zare1988}.
%, and $\theta$ and $\chi$ are the Euler angles
The effect of this interaction is to orient the electric dipole moment along the electric field direction,
which we characterize by $\expected{\cos\theta_{Z\mu}}$, with
$\theta_{Z\mu}$ being the angle formed by the permanent dipole moment with the LFF $Z$-axis.
In this work, we also explore the orientation of the MFF $z$ and $x$ axes along this LFF $Z$-axis 
in terms of $\expected{\cos\theta_{Zz}}$ and $\expected{\cos\theta_{Zx}}$, respectively. 

\subsection{Quantum Optimal Control}
\label{sec:quantum_optimal_control}

The goal of this work is to optimize the orientation of any molecular axis of an asymmetric planar molecule by using the Quantum Optimal Control (QOC) methodology
to design a time-dependent electric field while imposing certain restrictions.
%In this section we briefly describe the Quantum Optimal Control (QOC) methodology and how we apply it to the system under study. 
%The aim of QOC is to maximize (or minimize) a given expectation value at a certain time by means of an external field. 
To do so, we define the following functional~\cite{Werschnik2007} 
%$\mathcal{J}(\chi,\psi,E,\alpha)$,
%\rgf{which depends on a Lagrange multiplier $\chi(t)$, the wavefunction $\psi(t)$\footnote{Note that we omit any spatial dependency in the functions for the sake of clarity},
%the driving field $E(t)$, and a penalty factor $\alpha$}, 
%This functional is defined as
\begin{eqnarray}
\nonumber
\mathcal{J}(\chi,\psi,E,\alpha)&=&\mathcal{J}_\text{S}(\chi,E,\psi)+\mathcal{J}_\text{p}(E,\alpha)+\mathcal{J}_\text{o}(\psi)\\
\label{eq:functional}
&&
\end{eqnarray}
and seek for stationary trajectories in the system variables,~\ie, the driving field $E(t)$ and the wavefunction $\psi(t)$\footnote{Note that we omit any spatial dependency in the functions for the sake of clarity}. 
The first term in $\mathcal{J}(\chi,\psi,E,\alpha)$,
%where 
%\begin{itemize}
    %\item 
    %$\mathcal{J}_\text{S}(\chi,E,\psi)$
%the first term  
guarantees that the time-dependent Schr\"odinger equation is fulfilled by introducing the Lagrange multiplier $\chi(t)$
\begin{equation}
\label{eq:functional_schrodinger}
    \mathcal{J}_\text{S}(\chi,E,\psi)=-2\operatorname{Im}\left[\int_{T_0}^T\mathrm{d}t\melement{\chi(t)}{i\partial_t-H(t)}{\psi(t)}\right],
\end{equation} 
where $\operatorname{Im}$ denotes the imaginary part, and 
$T_0$ and $T$ are the initial propagation time and the measuring time, respectively. 
The penalty functional $\mathcal{J}_\text{p}(E,\alpha)$ reads as
\begin{equation}
\label{eq:functional_penalty}
    \mathcal{J}_\text{p}(E,\alpha)=-\alpha\int_{T_0}^T\mathrm{d}t\cfrac{E(t)^2}{S(t)},
\end{equation}
and forces a low fluence of the field by tuning the mask function $S(t)$ and a penalty factor $\alpha$. 
%\item 
The last term $\mathcal{J}_\text{o}(\psi)$ maximizes the expectation value of the operator $O$ at the final time $t=T$
\begin{equation}
    \label{eq:functional_observable}
    \mathcal{J}_\text{o}(\psi)=\melement{\psi(T)}{O}{\psi(T)}.
\end{equation}
%\end{itemize}
Note that the stationary trajectory fulfils that $\mathcal{J}_\text{S}(\chi,E,\psi)=0$ and maximizes (minimizes) $\mathcal{J}_\text{p}(E,\alpha)+\mathcal{J}_\text{o}(\psi)$.
To achieve the latter, the values of $\alpha$ and the function $S(t)$ must be carefully chosen such that the contribution of the fluence of $E(t)$ is small compared to  $\expected{O}$. 

The QOC equations~\cite{Werschnik2007} are obtained by solving the Euler equation associated to the functional $\mathcal{J}(\chi,\psi,E,\alpha)$ Eq.~\eqref{eq:functional}, and are given by
\begin{eqnarray}
    \label{eq:qoc_psi}
    &&i\partial_t\psi(t)-H(t)\psi(t)=0,\\
    \label{eq:qoc_chi}
    &&i\partial_t\chi(t)-H(t)\chi(t)=0,\\
    \label{eq:qoc_chi_psi}
    &&\chi(T)=O\psi(T),\\
    \label{eq:qoc_electric}
    && \vec{E}(t)=-\cfrac{S(t)}{\alpha}\operatorname{Im}\left[\melement{\chi(t)}{\vec{\mu}}{\psi(t)}\right],
\end{eqnarray}
and their complex conjugates. This self-consistent system of equations is solved using the Krotov method~\cite{Krotov1983,Tannor1992,Werschnik2007}, which converges to a stationary solution for positive defined observable operators~\cite{Ohtsuki2004}. 
Hence, to maximize the orientation $\expected{\cos\theta_{Zk}}$ and to ensure the convergence of the algorithm, we set $O=\cos\theta_{Zk}+1$.

In order to describe the dynamics, we solve the time dependent Schr\"odinger equation associated to the Hamiltonian in Eq.~\eqref{eq:hamiltonian}. 
A basis set expansion of the wavefunction $\Psi(\Omega, t)$ is done in terms of the Wigner matrix elements~\cite{Zare1988},
their most relevant properties and the matrix elements required are given in detailed elsewhere~\cite{Zare1988}.
In this basis expansion, we take into account that magnetic quantum number $M$ is conserved, because 
the electric field $\vec{E}(t)$ is parallel to the LFF $Z$-axis.
For the time propagation,  we apply the short iterative Lanczos scheme~\cite{Park1986}. 

\section{Results}
\label{sec:results}
\begin{figure}
    \centering
    \includegraphics[width=.5\linewidth]{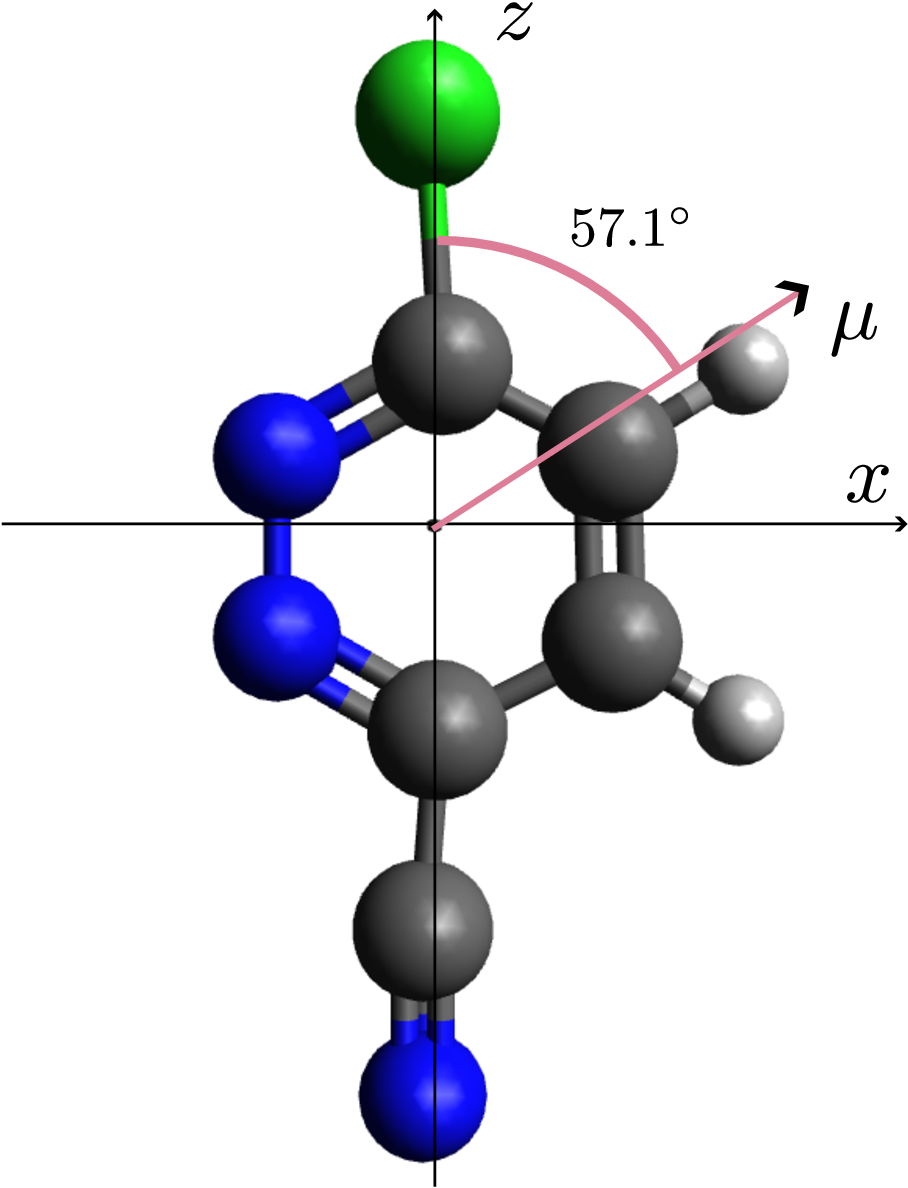}
    \caption{\label{fig:fig1} Sketch of the molecular structure of 6-chloropyridazine-3-carbonitrile (CPC). The molecular geometry and the dipole moment, $\vec\mu$, are taken from Ref.~\onlinecite{Hansen2013}.}
\end{figure}
As an example of non-symmetric planar molecules, we use 6-chloropyridazine-3-carbonitrile (CPC), with rotational constants $B_{x}=717.42$~MHz, $B_{y}=639.71$~MHz and $B_{z}=5905$~MHz~\cite{Hansen2013}.
This molecule is characterized by having its electric dipole moment not parallel to any principal axis of inertia, 
as shown in \autoref{fig:fig1}, with $\mu=5.20$~D and components $\mu_x=4.37$~D and $\mu_z=2.83$~D~\cite{Hansen2013}. %~\cite{Hansen2013,Thesing2017}. 
We focus on the field-dressed dynamics of the rotational ground state.
To provide a deeper physical insight into the QOC mechanism, we explore the orientation along the driving field direction of
any molecular axis, \ie, the permanent dipole moment, $\Vec{\mu}$, and the principal axes of inertia in the molecular plane $x$ and $z$.

For the mask function of the electric field, we use a Gaussian envelope $S(t)=\exp\left({-\cfrac{4t^2\ln{2}}{\tau^2}}\right)$, 
with $\tau$ being the Full Width at Half Maximum (FWHM).
The time interval is set to $[T_0,T]=[-2.53\tau,2.53\tau]$, taking  $T=-T_0=\sqrt{\cfrac{\ln{5\cdot 10^7}}{\ln 2}}\cfrac{\tau}{2}$ for computational convenience.
The field-dressed dynamics is analyzed in terms of transition between field-free rotational states involved in the optimal control process. The most relevant transitions and their frequencies are collected in \autoref{tab:spectrum}. 
These frequencies also allow us to interpret the structure of the driving field in the forthcoming sections.
\begin{table}[h]
\centering
\begin{tabular}{cccc}
\hline
Transition & Frequency  &Transition & Frequency \\
 &  (GHz) & &  (GHz) \\
\hline
$0_{00}0 \leftrightarrow 1_{01}0$ & 1.35 & $4_{04}0 \leftrightarrow 5_{05}0$ & 6.76 \\
$1_{10}0 \leftrightarrow 2_{02}0$ & 2.47  &$1_{01}0\leftrightarrow 2_{11}0$ & 7.81 \\
$1_{10}0 \leftrightarrow 2_{11}0$ & 2.63 & $2_{02}0 \leftrightarrow 3_{12}0$ & 9.05 \\
$1_{01}0 \leftrightarrow 2_{02}0$ & 2.71 & $2_{20}0 \leftrightarrow 3_{12}0$ & 11.87 \\ 
$2_{02}0 \leftrightarrow 3_{03}0$ & 4.06 & $8_{08}0\leftrightarrow 9_{18}0$ & 15.84 \\
$3_{03}0 \leftrightarrow 4_{04}0$ & 5.41 & $1_{10}0 \leftrightarrow 2_{20}0$ & 18.40 \\
$0_{00}0 \leftrightarrow 1_{10}0$ & 6.53 & $2_{20}0\leftrightarrow 3_{30}0$ & 30.27\\
\hline
\end{tabular}
\caption{\label{tab:spectrum} Transition frequencies between the main rotational states involved in the field-dressed dynamics.}
\end{table}

\subsection{Quantum Optimal Control for fixed penalty parameter $\alpha$}
\label{sec:qoc_for_fixed_alpha}
In this section, we set $\alpha=10^6$ for the driving electric field, which is equivalent 
to fixing its maximum allowed strength given by $\mu/\alpha$,
and consider two different FWHMs  $\tau$. 
\begin{figure*}
    \centering
\includegraphics[width=.95\linewidth]{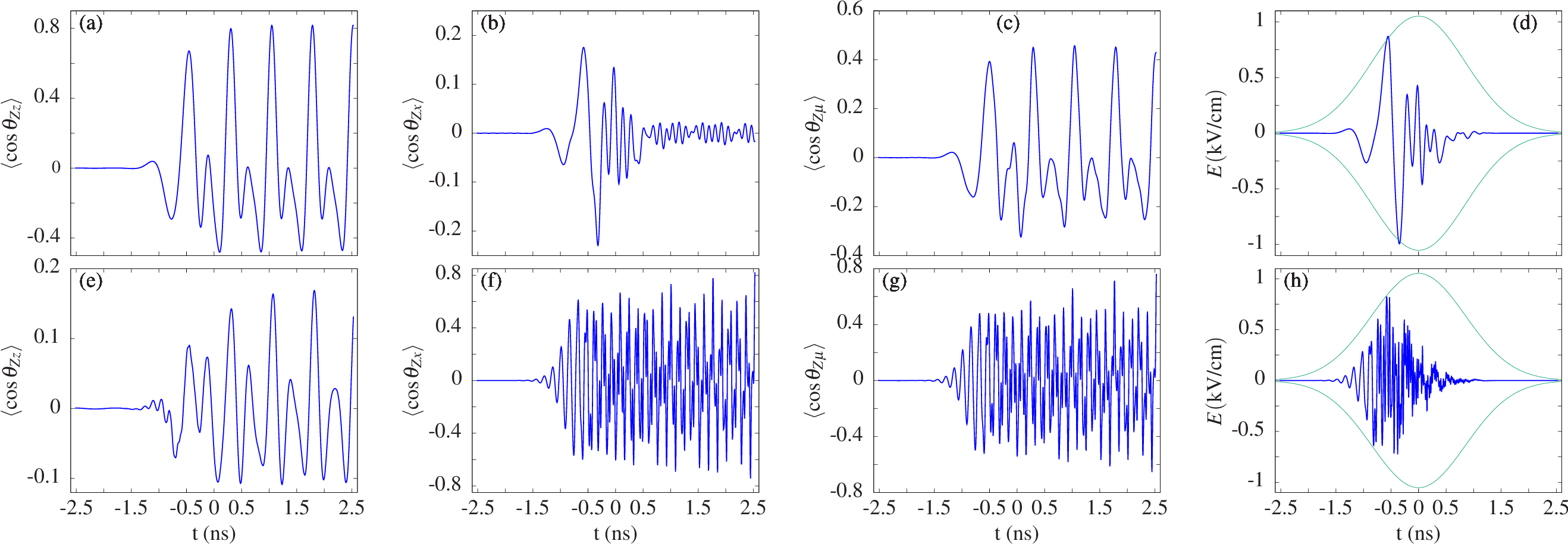}
    \caption{\label{fig:fig2}  
    Optimization of $\langle\cos\theta_{Zz}\rangle$ (panels (a)-(d) in the upper row) and $\langle\cos\theta_{Zx}\rangle$ 
    (panels (e)-(h) in the lower row) for the initial state $0_{00}0$. (a) and (e): $\langle\cos\theta_{Zz}\rangle$; (b) and (f): $\langle\cos\theta_{Zx}\rangle$; (c) and (g): $\langle\cos\theta_{Z\mu}\rangle$; and  (d) and (h): optimal electric field $E(t)$
    and mask function $\pm \frac{S(t)}{\alpha}$ as a function of time. 
    The parameters $\tau$ and $\alpha$ are set to $1$~ns and $10^6$, respectively.}
\end{figure*}
For  a $1$~ns-FWHM field, the results obtained when the  expectation value $\expected{\cos\theta_{Zz}}$ 
is optimized are presented in \autoref{fig:fig2}~(a)-(d).
The driving field designs a wavepacket populating rotational states and adjusting their relative phases. 
The time-evolution of the quantum interferences between these populated states provokes that 
$\expected{\cos\theta_{Zz}}$ reaches a maximum at the final $T$ as shown in \autoref{fig:fig2}~(a). 
During the time-evolution, the orientation varies between $ \expected{\cos\theta_{Zz}}=-0.48$  and
 $\expected{\cos\theta_{Zz}}=0.824$, and its oscillation pattern indicates that only a few states are involved in the 
 dynamics. A spectral analysis of $\expected{\cos\theta_{Zz}}$  encounters  three  main frequencies $1.38$, $2.77$ and $4.15$~GHz, associated 
 to the transitions $0_{00}0\leftrightarrow 1_{01}0$, $1_{01}0\leftrightarrow 2_{02}0$ and $2_{02}0\leftrightarrow 3_{03}0$, respectively, see \autoref{tab:spectrum}. 
 These rotational states are linked by the selection rules 
 $\Delta J=\pm 1$ and $\Delta M=0$, imposed by the operator $\cos\theta_{Zz}$~\cite{Zare1988,Omiste2011a} in~\autoref{eq:efield}.
For this optimization, the driving electric field in \autoref{fig:fig2}~(d) is mostly relevant before $t=0$ finding two main peaks at 
 $t=-0.56$~ns and $t=-0.35$~ns with intensity $0.87$~kV/cm and $-1$~kV/cm, respectively, being very weak afterwards. 
This control field is also mainly formed by these transition frequencies, as illustrated by the square of its
Fourier Transform  $|\mathcal{F}[E(t)]|^2$  in \autoref{fig:fig3}. 
Note that $|\mathcal{F}[E(t)]|^2$ is normalized so its maximum value is $1$.
The wide peaks of the $\tau=1$~ns field are due to the duration imposed in the optimization algorithm. One can distinguish the frequencies 
$1.35$, $4.06$ and $5.41$~GHz, associated to the transitions $0_{00}0 \leftrightarrow 1_{01}0$, $2_{02}0 \leftrightarrow 3_{03}0$ and 
$3_{03}0 \leftrightarrow 4_{04}0$, respectively.
Note that the contribution of the latter in the orientation pattern is negligible.
The frequency $2.77$~GHz from the $1_{01}0 \leftrightarrow 2_{02}0$ transition is  blurred due to the overlap with the preceding peaks. 
 \begin{figure}[b]
    \centering
    \includegraphics[width=1\linewidth]{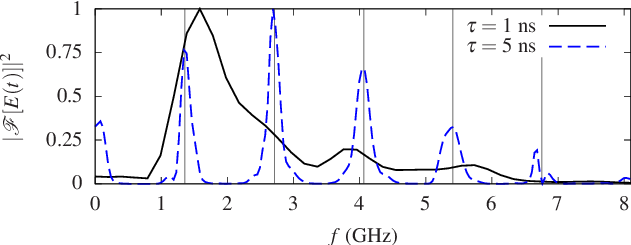}
    \caption{\label{fig:fig3}  
    For $\alpha=10^6$ and FHWMs $\tau=1$ and $5$~ns, the absolute squared Fourier Transform of the optimized electric field, 
    $|\mathcal{F}[E(t)]|^2$ to maximize $\expected{\cos\theta_{Zz}}$. Note that $\left| \mathcal{F}[E(t)]\right|^2$ is normalized
    so that the absolute maximum equals $1$.}
\end{figure}
Despite the field being designed to optimize $\expected{\cos\theta_{Zz}}$, we encounter 
a moderate orientation of the molecular $x$-axis in \autoref{fig:fig2}~(c) due to the coupling 
through $\mu_x$. 
The orientation $\expected{\cos\theta_{Zx}}$ faithfully follows the external field until $t\approx 0.5$~ns.
From there on the field is very small, and $\expected{\cos\theta_{Zx}}$ oscillates with a small amplitude
reaching $\expected{\cos\theta_{Zx}}=-0.028$ at $t=T$.
Reaching this almost zero orientation is possible since the time scale required for the $x$-axis dynamics is shorter than the period of oscillation of the electric field. Hence the impact of the driving field averages out due to the rapid oscillation of the molecule. The frequency of the $x$ axis orientation is determined by the rotational constants of the orthogonal axes, namely 
$B_y$ and $B_z$, which have an average value of approximately $3.27$~GHz. This frequency is greater than the oscillation frequency of 
the driving field, which is around $1.6$~GHz.
The spectral analysis of $\expected{\cos\theta_{Zx}}$ confirms that the 
states $1_{10}0$, $2_{11}0$ and $3_{12}0$ are also populated in the field-dressed dynamics.
For the sake of completeness, we present $\expected{\cos\theta_{Z\mu}}$ in \autoref{fig:fig2}(c), which shows a similar behaviour as $\expected{\cos\theta_{Zz}}$.

The driving field and the dressed rotational dynamics are highly complex when the QOC technique is applied to the orientation of the 
molecular $x$-axis.
In \autoref{fig:fig2}~(f), $\expected{\cos\theta_{Zx}}$ shows a complex oscillatory behaviour,
which is due to the different time scale associated with the dynamics along this axis.
At $t=T$ a large orientation is reached $\expected{\cos\theta_{Zx}}\approx 0.83$.
The driving field is presented in \autoref{fig:fig2}~(h), 
$E(t)$ rapidly oscillates for $t\lesssim 0.5$~ns, which contrasts with the slow oscillations of 
field optimizing $\expected{\cos\theta_{Zz}}$ in \autoref{fig:fig2}~(d).
This behaviour agrees with the rapid rotational dynamics of the MFF $x$-axis indicated above, and  
$\expected{\cos\theta_{Zx}}$ follows the field oscillations in \autoref{fig:fig2}~(f). 
The Fourier Transform of $\expected{\cos\theta_{Zx}}$ shows three main frequencies, namely, $6.45$~GHz, $18.47$~GHz 
and $30.40$~GHz, associated to the transitions $0_{00}0\leftrightarrow 1_{10}0$, $1_{10}0\leftrightarrow 2_{20}0$ and 
$2_{20}0\leftrightarrow 3_{30}0$, respectively.
Now, the orientation of the dipole moment along the LFF $Z$-axis $\expected{\cos\theta_{Z\mu}}$,
see \autoref{fig:fig2}~(g) follows a similar behaviour as $\expected{\cos\theta_{Zx}}$, and a significant value
is reached at $t=T$ with $\expected{\cos\theta_{Z\mu}}=0.76$.

In contrast,  $\expected{\cos\theta_{Zz}}$  is incapable of adjusting to the rapidly changing driving field as observed in \autoref{fig:fig2}~(e).  
Indeed, the control field in \autoref{fig:fig2}~(h) imprints short transfers of momentum at each cycle, similar to the impulsive alignment described for short laser pulses~\cite{stapelfeldt:rev_mod_phys_75_543}. 
As a consequence,  the amplitude of $\expected{\cos\theta_{Zz}}$ cannot be efficiently reduced at the final time in general.
To bypass this effect, the algorithm fine tunes all the relative quantum phases within the wavepacket, seeking for 
reducing the orientation of the $z$-axis, \ie, approaching it to zero,  at the final time $t=T$. 
However, if the pulse duration is short compared to the time characteristic of the
 $z$-axis dynamics, the oscillation may not vanish. This is the case in \autoref{fig:fig2}~(e), where the 
 orientation of the molecular $z$-axis is not negligible, \ie, $\expected{\cos\theta_{Zz}}\approx 0.16$, at $t=T$, 
 which is due to the short FWHM, $\tau=1$~ns, of the driving field. 
\begin{figure*}
    \centering
    \includegraphics[width=.95\linewidth]{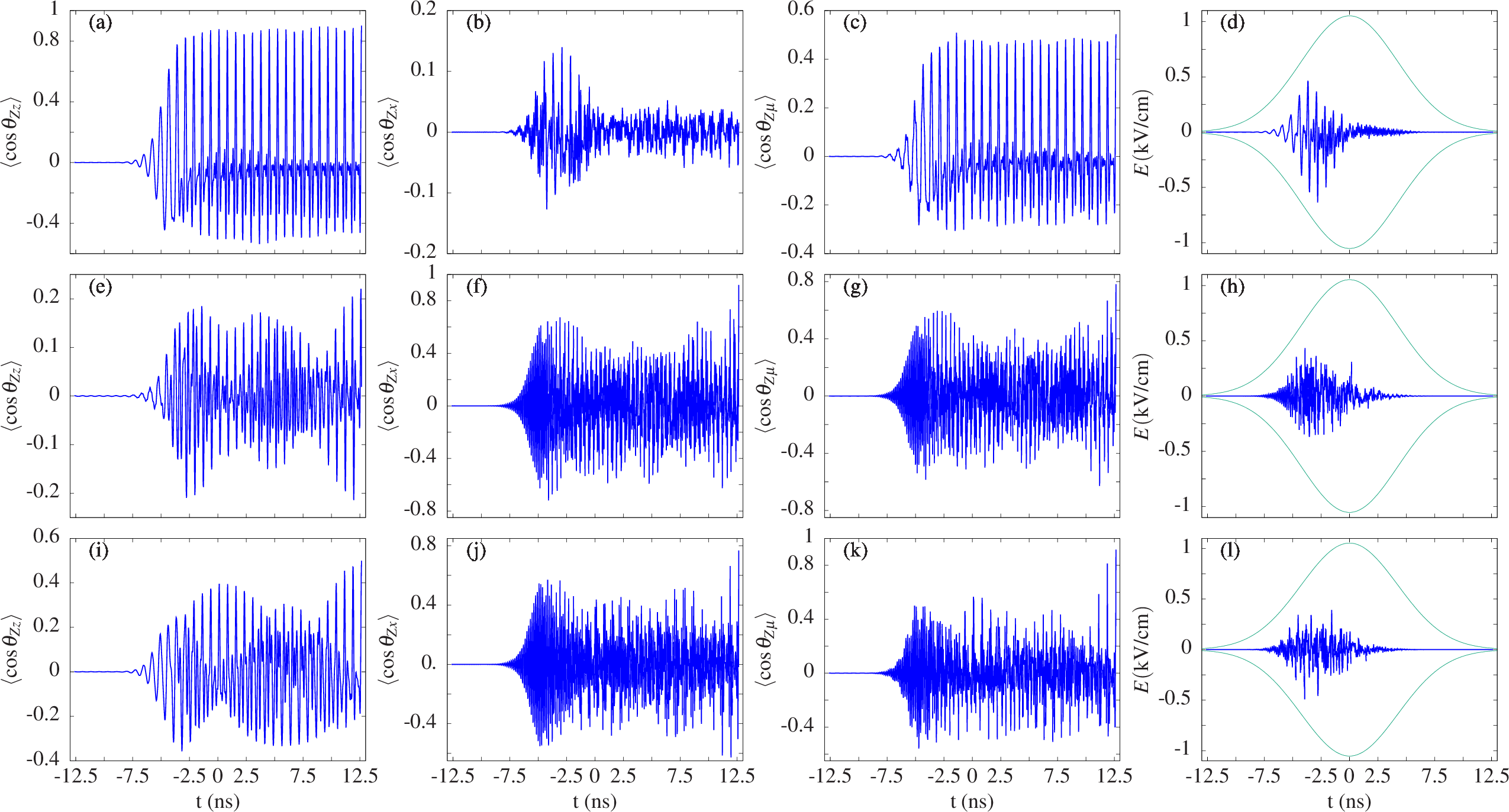}
\caption{\label{fig:fig4} Optimization of $\langle\cos\theta_{Zz}\rangle$ (panels (a)-(d) in the upper row) and 
$\langle\cos\theta_{Zx}\rangle$ (panels (e)-(h) in the middle row) and $\langle\cos\theta_{Z\mu}\rangle$ 
(panels (e)-(h) in the lower row) for  the initial state $0_{00}0$. 
(a), (e) and (i): $\langle\cos\theta_{Zz}\rangle$; (b), (f) and (j): $\langle\cos\theta_{Zx}\rangle$; 
(c), (g) and (k): $\langle\cos\theta_{Z\mu}\rangle$; 
and (d), (h) and (l): optimal electric field $E(t)$
    and mask function $\pm \frac{S(t)}{\alpha}$  as a function of time. 
    The parameters $\tau$ and $\alpha$ are set to $5$~ns and $10^6$, respectively.}    
\end{figure*}

We now consider an optimizing field with $\tau=5$~ns FWHM, and explore the results of the QOC algorithm applied to the
three possible orientation axes.
Compared to the $1$~ns pulse in \autoref{fig:fig2}~(a), 
$\expected{\cos\theta_{Zz}}$ exhibits faster oscillations with large amplitude, which 
increase till the maximal value $\expected{\cos\theta_{Zz}}=0.9$ at $t=T$. 
The optimized electric field, see \autoref{fig:fig4}~(a), also possesses  rapid oscillations
mainly for $t\lesssim 0$~ns when the rotational wavepacket is created.
In particular, for  $t\lesssim -3.5$~ns,  $\expected{\cos\theta_{Zz}}$ follows this driving field.
The spectral analysis given by the Fourier Transform of the field 
in \autoref{fig:fig3} presents  clearly defined and well separated peaks.
This larger FWHM allows to populate highly excited rotational states associated to
the additional frequency $6.76$~GHz of the $4_{04}0\leftrightarrow 5_{05}0$  transition.
Panels (b) and (c) of \autoref{fig:fig4} present the orientation of the MFF $x$ axis and $\vec{\mu}$
along the LFF $Z$-axis.
As in the previous case, the behaviour of $\expected{\cos\theta_{Z\mu}}$ resembles $\expected{\cos\theta_{Zz}}$, but
reaching smaller final value with $\expected{\cos\theta_{Z\mu}}=0.502$ at $t=T$.
In contrast, $\expected{\cos\theta_{Zx}}$ shows rapid oscillations of  small amplitude for $t>0$~ns, and 
at $t=T$ $\expected{\cos\theta_{Zx}}=0.015$.

The results obtained when $\expected{\cos\theta_{Zx}}$ is optimized are shown in \autoref{fig:fig4}~(e)-(h).
The driving field  in \autoref{fig:fig4}~(h), presents a more complex
structure with faster and narrower oscillations compared to the previous field in \autoref{fig:fig4}~(d).
The number of rotational states contributing to the field-dressed dynamics is enhanced, as it is manifested in the irregular
oscillations of $\expected{\cos\theta_{Zz}}$, $\expected{\cos\theta_{Zx}}$ and $\expected{\cos\theta_{Z\mu}}$
in \autoref{fig:fig4}~(e),~(f) and~(g), respectively. 
During the time evolution, $\expected{\cos\theta_{Zz}}$ shows moderate values, but the field is adjusted to reduce 
it at $t=T$ attaining $\expected{\cos\theta_{Zz}}\approx 0.019$. 
In contrast, the orientation  of the MFF $x$-axis is significantly enhanced to $\expected{\cos\theta_{Zx}}\approx 0.92$
at $t=T$, and $\expected{\cos\theta_{Z\mu}}$ presents a similar behaviour. 

For completeness, we present the optimization of the orientation of the dipole-moment axis in \autoref{fig:fig4}~(i)-(l). In this case, the quantum interference is constructive for $\expected{\cos\theta_{Zx}}$ and $\expected{\cos\theta_{Zz}}$, having both a local maximum at $t=T$
$\expected{\cos\theta_{Zz}}\approx 0.50$ and $\expected{\cos\theta_{Zx}}\approx 0.77$,
whereas  $\expected{\cos\theta_{Z\mu}}\approx 0.92$. This illustrates that the orientation of the permanent dipole moment is 
largely dominated by its $x$ component. The spectral analysis of $E(t)$ shows that the transitions collected in \autoref{tab:spectrum} are relevant for the QOC, \ie, these rotational states build up the control dynamics and 
maximal orientation of $\vec{\mu}$.

\subsection{Quantum Optimal Control for varying the penalty factor $\alpha$}
\label{sec:qoc_for_varying_alpha}
\begin{figure}[b]
    \centering
    \includegraphics[width=.95\linewidth]{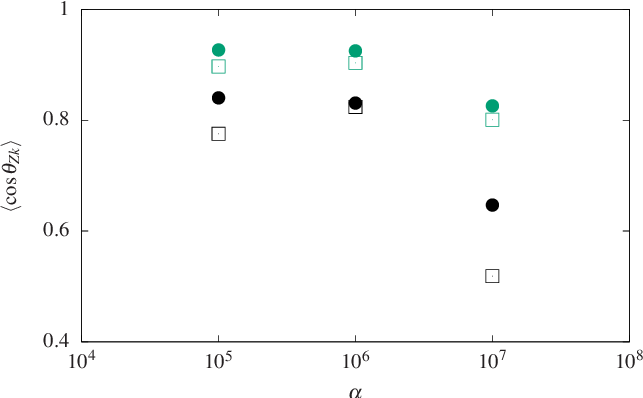}
    \caption{\label{fig:fig5} 
    Final orientations of the MFF $x$ (bullets) and $z$ (squares) axes,
  $\expected{\cos{\theta_{Zx}}}$   and $\expected{\cos{\theta_{Zz}}}$, respectively,  
    after applying QOC 
    using electric fields with FWHM $\tau=1$ (black) and $5$~ns (green)
    as a function of the penalty factor $\alpha$.}
\end{figure}

In this section, we investigate the QOC orientation for different values of $\alpha$, \ie, 
the maximum allowed strength of the control field, which is represented by the factor $\mu/\alpha$. 
For $\tau=1$~and $5$~ns, we present  in \autoref{fig:fig5} the final optimized orientation 
as a function of the penalty factor $\alpha$.
Our analysis reveals that for a given $\alpha$ and $\tau$, optimizing $\expected{\cos\theta_{Zx}}$ results in a larger orientation compared to $\expected{\cos\theta_{Zz}}$. This finding supports the idea that QOC is more effective when the natural timescale of the degree of freedom being optimized is smaller than the field duration, as discussed in \autoref{sec:qoc_for_fixed_alpha}. Moreover, 
for a given $\alpha$, the larger the FWHM the better the optimized orientation is, for both $\expected{\cos\theta_{Zx}}$ and $\expected{\cos\theta_{Zz}}$.

Our analysis also reveals that, for the values of $\alpha$ considered, $\expected{\cos\theta_{Zx}}$ decreases as a function of $\alpha$, 
meaning that stronger fields result in larger optimized orientations.  
Note that for both FWHMs, $\expected{\cos\theta_{Zx}}$ is very similar for $\alpha=10^5$ and $10^6$,
\ie, the orientation saturates beyond certain intensities of the driving field. Since the orientation is not increasing, the functional $\mathcal{J}_\text{o}+\mathcal{J}_\text{p}$ is maximized by minimizing the fluence of the field, involved in the penalty functional, $\mathcal{J}_\text{p}$. 
This saturation of  $\expected{\cos\theta_{Zx}}$ with $\alpha$ can be explained in terms of the brute force orientation occurring at these
strong fields, combined with the larger frequencies along the MFF $x$-axis facilitating that the molecule follows the driving field.
Furthermore, for $\alpha=10^5$, that is, the stronger field case, the population transfer may not be understood in terms of one-photon transition, but as a strong field process.

In contrast,  the orientation of the MFF $z$-axis  decreases as $\alpha$ decreases, \ie, as the maximal field strength increases.
This is due to the strong dependence of the controllability of the orientation of this axis on the process for the suppression of $\expected{\cos\theta_{Zx}}$ imposed by the QOC equations. 
Note that the faster dynamics of the $x$ axis and the larger component of $\Vec{\mu}$ along $x$ also play an important role against reducing $\expected{\cos\theta_{Zx}}$. 
Thus, achieving this suppression requires finely tuned quantum phases among the rotational states, as small changes can result in different outcomes. Besides, the QOC algorithm aims to maximize $\mathcal{J}_\text{p}(E,\alpha)+\mathcal{J}_\text{o}(\psi)$, as discussed in \autoref{sec:quantum_optimal_control}. However, due to the limited control of $\mathcal{J}_\text{o}(\psi)$, a better approach to the stationary point is reached for smaller $\mathcal{J}_\text{o}(\psi)$ and larger $\mathcal{J}_\text{p}(E,\alpha)$.

\begin{figure}
    \centering
    \includegraphics[width=\linewidth]{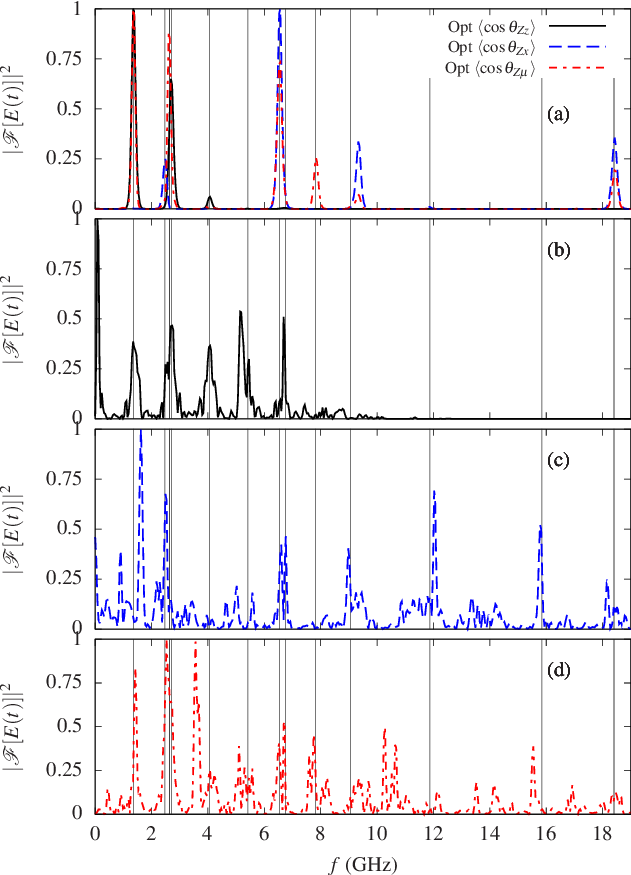}
    \caption{\label{fig:fig6} For a pulse duration of $\tau=5$ns, the absolute squared Fourier Transform, $\left| \mathcal{F}[E(t)]\right|^2$, of the optimized electric field is shown for (a) $\alpha=10^7$ and (b)-(d) $\alpha=10^5$. 
    The Fourier Transform plotted correspond to optimization of $\langle\cos\theta_{Zz}\rangle$ (solid black), $\langle\cos\theta_{Zx}\rangle$ (blue dashed), 
    and $\langle\cos\theta_{Z\mu}\rangle$ (red dot-dashed) and taking as initial state $0_{00}0$. 
    The vertical lines correspond to the transition frequencies shown in \autoref{tab:spectrum}. Note $\left| \mathcal{F}[E(t)]\right|^2$ is normalized so that the absolute maximum equals 1.}
\end{figure}

Finally, we perform an spectral analysis of the  optimization along the $x$, $z$, and $\mu$ axes for the impact parameter $\alpha=10^5$ and $10^7$ and $\tau=5$~ns.
The Fourier transforms of the QOC fields are shown in \autoref{fig:fig6}, together with the frequencies of the main transitions between field-free 
rotational states, which are also collected in \autoref{tab:spectrum}.
In the weak field regime, $\alpha=10^7$, shown in \autoref{fig:fig6}~(a), we observe well-defined isolated frequencies for all the optimized orientation observables. 
The frequencies of  $E(t)$ obtained from the optimization of $\expected{\cos\theta_{Zz}}$ and $\expected{\cos\theta_{Zx}}$ are different due to the different selection rules 
$\Delta K=0$ and  $\Delta K=\pm 1$ imposed by the operators $\mu_z$ and $\mu_x$, respectively.
For the initial state $0_{00}0$, the relevant transitions for the optimization of the $z$-axis orientation are $0_{00}0\leftrightarrow 1_{01}0$, 
$1_{01}0\leftrightarrow 2_{02}0$, and $2_{02}0\leftrightarrow 3_{03}0$, which are characterized by frequencies of $1.35$, $2.71$, and $4.06$~GHz, respectively. 
However, the optimization of $\expected{\cos\theta_{Zx}}$ is dominated by the frequencies mixing $K$, specifically $0_{00}0\leftrightarrow 1_{10}0$ ($6.53$~GHz), 
$1_{10}0\leftrightarrow 2_{02}0$ ($2.47$~GHz), $2_{02}0\leftrightarrow 3_{12}0$ ($9.05$~GHz). 
The relatively low value of $\expected{\cos\theta_{Zx}}$ at $t=T$, see~\autoref{fig:fig5},
 can be explained in terms of the relevant frequencies in $E(t)$ for the orientation of $\expected{\cos\theta_{Zz}}$ being below the frequency required to transfer population from $0_{00}0$ to $1_{10}0$, which 
is the first excited state coupled by $\mu_x$. On the contrary, the transition frequencies of $0_{00}0\leftrightarrow 1_{10}0$ and $2_{02}0\leftrightarrow 3_{12}0$ are above the main transitions.
This results in the impulsive orientation of the molecular $Z$-axis 
as discussed in \autoref{sec:qoc_for_fixed_alpha}.
Additional frequencies are observed  in \autoref{fig:fig6}~(a) when the orientation of the permanent dipole moment is optimized,
with two new peaks at $2.63$ and $7.81$~GHz due to $1_{10}0\leftrightarrow 2_{11}0$ and $1_{01}0\leftrightarrow 2_{11}0$, respectively.

For a strong control field with $\alpha=10^5$, the spectrum becomes more complex due to two main reasons. 
Firstly, the contribution of more excited rotational states, such as the peak at $15.84$GHz corresponding to the transition $8_{08}0\leftrightarrow 9_{18}0$, 
which appears for the orientation of the MFF $x$-axis along the LFF $Z$-axis. Secondly, the brute force orientation mediated by the large amplitude of the field, as reported in previous studies\cite{loesch:jcp93,bulthuis:jpca101}. Therefore, the driving field can not be simply described as the contribution of several frequencies,~\ie, as population transfer mediated by absorption of several photons, due to the high intensity of the driving field, as we discussed in \autoref{sec:qoc_for_varying_alpha}.

\section{Conclusions and outlook}
\label{sec:conclusions_and_outlook}

In this work, we have demonstrated the feasibility of fully controlling the orientation of any molecular axis
of non-symmetric planar molecules by means of quantum optimal control. 
The driving field is composed of just a few frequencies corresponding to transitions among several field-free states, and its 
temporal profile and strength can be reached experimentally.
For a mask function with  FWHM and weak strength, 
our findings show limited control due to the small population transfer stimulated due to the interaction with the field. 
On the contrary, the electric field cannot be described by isolated components for strengths above $1$~kV/cm, since 
 the number of transitions among the involved states increases and they overlap due to their spectral width. 
 This effect is further enhanced for short pulses. 
This work shows the efficiency, flexibility and potential of QOC for the  rotational dynamics control of non-symmetric molecules, and
provides valuable insights into the conditions required for it.

This study focus on the stereodynamics of the ground state of CPC, and without loss of generality, the results could be extended
to excited states.
This technique provides an alternative strategy to the mixed-field orientation method, which, in the case of CPC,  
was shown to attain full control of only one molecular axis~\cite{Hansen2013,Thesing2017}. 

Summing up, we have demonstrated that QOC is a powerful technique to control the orientation of any planar molecule without rotational symmetry, and,
 furthermore, 
to more complex systems as enantiomers. Indeed, the manipulation, control and selection of chiral 
molecules~\cite{Leibscher2022a,Goerz2019, Leibscher2023}
will have a enormous impact on the pharmaceutical industry, where the spatial arrangement of the atoms in one enantiomer may determine the biological activity of a drug, as the remarkable case of thalidomide~\cite{Eriksson1995}.
Moreover, the design of optimally designed chiral electric fields could be used to efficiently create oriented superrotor states~\cite{Tutunnikov2021a}. 

\begin{acknowledgments}
J.J.O. acknowledges the funding by the Madrid Government (Comunidad de Madrid Spain) under the Multiannual 
Agreement with Universidad Complutense de Madrid in the
line Research Incentive for Young PhDs, in the context of the V PRICIT (Regional Programme of Research and Technological Innovation) 
(Grant: PR27/21-010), spanish projects PID2019-105458RB-I00 and PID2021-122839NB-I00 (MICIN). 
R.G.F. gratefully acknowledges financial support by the Spanish projects PID2020-113390GB-I00 (MICIN),
PY20$\_$00082 (Junta de Andalucía), and A-FQM-52-UGR20 (ERDF-University of Granada) 
and the Andalusian research group FQM-207.
We would also like to thank Ignacio Solá for fruitful discussions.  
\end{acknowledgments}

%\bibliography{quantum_control_and_information_stereo}
%

\end{document}